\documentstyle[aps,pra,floats,twocolumn,epsfig]{revtex}

\begin{document}
\draft
\wideabs{
\title{Observations of a doubly driven V system probed to a fourth level in
laser-cooled rubidium}
\author{S. R. de Echaniz$^{1}$, Andrew D. Greentree$^{1}$, A. V. Durrant$^{1}$, D.
M. Segal$^{2}$, J. P. Marangos$^{2}$ and J. A. Vaccaro$^{3}$}
\address{$^{1}$Quantum Processes Group, Department of Physics and Astronomy, The Open
University,\\
Walton Hall, Milton Keynes, MK7 6AA, UK\\
$^{2}$Laser Optics and Spectroscopy Group, Blackett Laboratory, Imperial
College of Science, Technology and Medicine,\\
London SW7 2BZ, UK\\
$^{3}$Department of Physics and Astronomy, University of Hertfordshire,\\
Hatfield AL10 9AB, UK}
\date{\today}
\maketitle

\begin{abstract}
Observations of a doubly driven V system probed to a fourth level in an N
configuration are reported. A dressed state analysis is also presented. \
The expected three-peak spectrum is explored in a cold rubidium sample in a
magneto-optic trap. Good agreement is found between the dressed state theory
and the experimental spectra once light shifts and uncoupled absorptions in
the rubidium system are taken into account.
\end{abstract}
}

\section{Introduction}

The optical properties of coherently prepared atoms have been much studied
in recent years. The simplest case is when one transition is resonantly
pumped with a strong coherent laser beam and the resulting dressed system
probed from a third level in a $\Lambda $, V or cascade configuration. These
systems can exhibit electromagnetically-induced transparency (EIT) where
quantum interference effects allow a resonant probe beam to propagate
without absorption and with greatly-enhanced dispersion \cite{bib:EIT}. The
EIT effect has been exploited in a variety of applications: gain without
inversion \cite{bib:GWI}, ground state cooling of trapped atoms\cite
{bib:Cooling}, ultra-slow light-speed propagation \cite{bib:UltraSlow},
quantum non-demolition measurements \cite{bib:QND} and efficient non-linear
optical processes \cite{bib:NLO}. This has lead to interest in more complex
systems involving multiple electromagnetic fields interacting with either a
single transition \cite{bib:PolyChro} or with several transitions \cite
{bib:Sandhya1997}\cite{bib:Sadeghi1997}\cite{bib:Wei1998} \cite{bib:Q4WM}.
In particular there is now huge interest in exploiting non-linearities in
four-level systems with applications in quantum optics \cite{bib:QNLO} and
four-wave mixing \cite{bib:Q4WM}.

Four-level systems of various configurations can be excited with different
pumping schemes. Several theoretical studies have considered three-level
ladder configurations coherently prepared by two strong laser beams \cite
{bib:Sandhya1997}\cite{bib:Sadeghi1997}\cite{bib:Wei1998}. They predict a
three-peak spectrum when the lowest or highest of the three levels is probed
via a fourth level or monitored by fluorescence. When the two dressing beams
are on resonance this spectrum has the form of a Doppler-free three-photon
absorption peak situated centrally within an EIT window \cite
{bib:Sandhya1997}. More generally, the positions and intensities of
the three peaks are functions of the Rabi frequencies and detunings of the
two strong beams. The trajectories of the peaks as a function of Rabi
frequencies and detunings have been described in doubly dressed analyses 
\cite{bib:Sadeghi1997}\cite{bib:Wei1998}. None of these three studies was
accompanied by experimental verifications.

In this paper we consider a similar scheme to those of references \cite
{bib:Sandhya1997}\cite{bib:Sadeghi1997}\cite{bib:Wei1998} but with the two
strong pump beams, or coupling beams as we shall call them, connecting a
ground or metastable level to two excited levels. The resulting dressed V
system can then be probed from another ground or metastable level, as
illustrated in Figure 1(a). An experimental realisation in $^{87}$Rb cooled
in a magneto-optic trap (MOT) is indicated in Figure 1(b). Our interest in
this system stems from our current programme of work on the strong cross-
and self-phase modulation expected in the N-configuration of Figure 1, and
the main aim of this paper is to characterise the weak-probe spectrum of
this system. For this task we have to be concerned not only with the
behaviour of the ideal four-level model (Figure 1(a)) but also with light
shifts due to off-resonant interactions with other nearby hyperfine levels
and the effects of level degeneracies in the $^{87}$Rb system. Because our
experiments are carried out in a Rb MOT, we are free to propagate our beams
in any desired directions, and to choose coupling beams on different optical
transitions that are well-separated in frequency thus avoiding potentially
complex mutual light shifts induced by the two beams, although some simple
light shifts and the effects of the trapping beams do have to be taken into
account in the system of Figure 1(a).

\begin{figure}[!tb]
\epsfig{figure=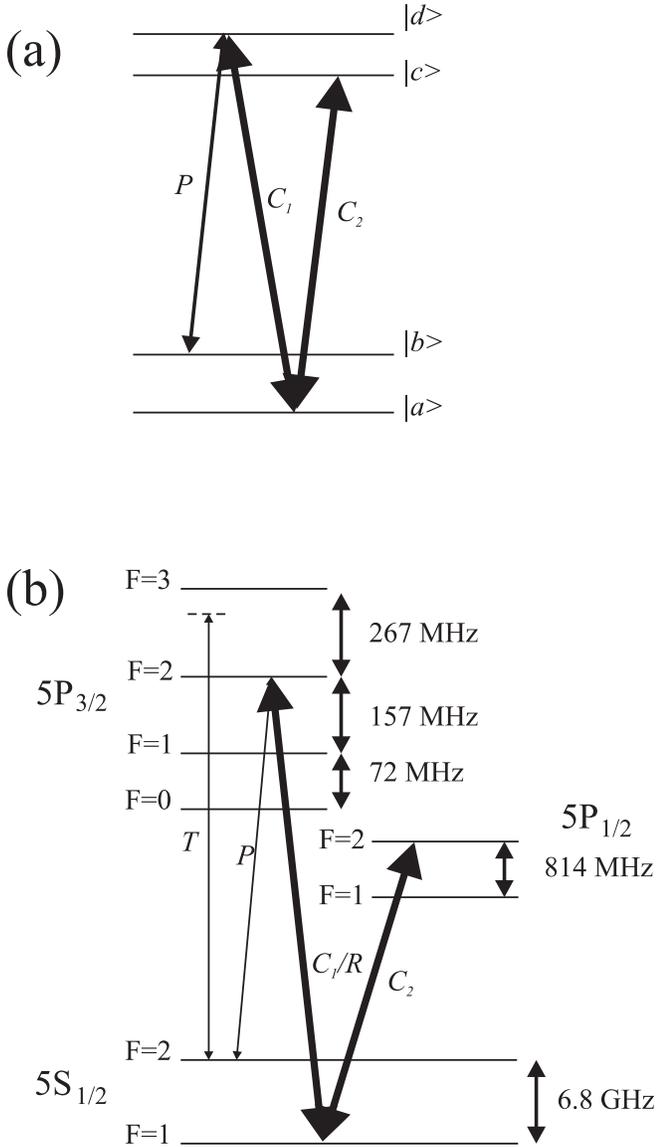,width=86mm}
\\
\caption{(a) The model four-state N configuration. $C_{1}$ and $C_{2}$ are strong
coupling fields driving a V system. $P$ is a weak probe field. (b) The real
$^{87}$Rb system in which the experiments are carried out. The atoms are
trapped and cooled in a magneto-optic trap by trapping fields T, detuned by
$-13\mathop{\rm MHz}$ from the $5S_{1/2}F=2$ to $5P_{3/2}F=3$ transition of
the D$_{2}$ line. (The coupling field $C_{1}$and the repumping field $R$ for
the MOT are derived from the same laser.)}
\end{figure}

Many manifestations of Zeeman degeneracy have been reported in EIT
experiments: optical pumping among Zeeman levels and to other levels \cite
{bib:ZeemanPumping}, coherent population trapping by the coupling beam \cite
{bib:CPT}, the inversion of EIT dips (i.e. electromagnetically induced
absorption) \cite{bib:EIA}, and absorption of the probe beam on Zeeman
transitions that are not coupled by the pump beam\cite
{bib:UncoupledAbsorption}. These latter uncoupled absorptions are a constant
feature of our probe spectra, superimposed on the V system spectra we are
interested in.

This paper is organised as follows. Section II presents a dressed state
analysis of the strongly-pumped model V system of Figure 1(a), showing the
expected dependance of the probe absorption spectrum on Rabi frequencies and
detunings. Section III describes the experimental arrangement and presents
EIT spectra with the first coupling field $C_{1}$ applied and the second
coupling field $C_{2}$ switched off, thus showing the effects of light
shifts and uncoupled absorptions in a relatively simple system. Our main
results are presented in Section IV where both coupling fields are applied
and probe spectra obtained for various coupling field intensities and
detunings. The spectra are interpreted using the 4-state theory of Section
II modified phenomenologically by the light shift and uncoupled absorption
effects described in Section III. We justify the use of the four-state
model and we describe the origin of the uncoupled absorptions relevant to
our experiments in the Appendix.

\section{Theory}

Figure 1(a) shows a four state N configuration, although for the purposes of
this discussion we shall consider it as a three state V scheme formed by the
states $\left| a\right\rangle $, $\left| c\right\rangle $ and $\left|
d\right\rangle $ and strong coupling fields $C_{1}$ and $C_{2}$ of
frequencies $\omega _{1}$ and $\omega _{2}$, probed weakly on the $\left|
b\right\rangle -\left| d\right\rangle $ transition. The Hamiltonian
describing the system consisting of the atom and the two coupling fields can
be written down in the semiclassical and rotating wave approximations as 
\[
{\cal H}={\cal A}+{\cal I} 
\]
where 
\begin{eqnarray*}
{\cal A} &=&\Delta _{1}\left| a\right\rangle \left\langle a\right| +\left(
\Delta _{1}-\Delta _{2}\right) \left| c\right\rangle \left\langle c\right| \\
{\cal I} &=&\frac{\Omega _{1}}{2}\left( \left| d\right\rangle \left\langle
a\right| +\left| a\right\rangle \left\langle d\right| \right) +\frac{\Omega
_{2}}{2}\left( \left| c\right\rangle \left\langle a\right| +\left|
a\right\rangle \left\langle c\right| \right) .
\end{eqnarray*}
${\cal A}$ and ${\cal I}$ represent the atomic and interaction parts of the
Hamiltonian ${\cal H}$. $\Delta _{1}=\omega _{1}-\omega _{da}$ ($\Delta
_{2}=\omega _{2}-\omega _{ca}$) is the detuning of coupling field $C_{1}$ ($%
C_{2}$) from the $\left| a\right\rangle -\left| d\right\rangle $ ($\left|
a\right\rangle -\left| c\right\rangle $) transition and $\omega _{\beta
\alpha }$ is the transition frequency of the $\left| \alpha \right\rangle
-\left| \beta \right\rangle $ transition for $\alpha ,\beta =a,b,c,d$. $%
\Omega _{j}={\bf d}_{j}\cdot {\bf E}_{j}$ is the Rabi frequency for field $%
j=1,2$ where the field ${\bf E}_{j}$ interacts only with its quasi-resonant
transition with electric dipole moment ${\bf d}_{j}$. We have chosen units
such that $\hbar =1$ so energies are measured in units of frequency.

Because there are three basis states which describe the system, in the most
general case, we expect a characteristic three line spectrum when performing
probe absorption experiments. Writing out the Hamiltonian in matrix form
gives 
\begin{equation}
{\cal H}=\left[ 
\begin{array}{lll}
\Delta _{1} & \Omega _{2}/2 & \Omega _{1}/2 \\ 
\Omega _{2}/2 & \Delta _{1}-\Delta _{2} & 0 \\ 
\Omega _{1}/2 & 0 & 0
\end{array}
\right] .  \label{eq:HamiltonianMatrix}
\end{equation}

Following the method in Shore \cite{bib:Shore1990} we derive the doubly
dressed state energies 
\begin{eqnarray*}
{\cal E}_{1} &=&-\frac{1}{3}\alpha +\frac{2}{3}p\cos \left( \frac{\Theta }{3}%
\right) \\
{\cal E}_{2} &=&-\frac{1}{3}\alpha -\frac{2}{3}p\cos \left( \frac{\Theta }{3}%
+\frac{\pi }{3}\right) \\
{\cal E}_{3} &=&-\frac{1}{3}\alpha -\frac{2}{3}p\cos \left( \frac{\Theta }{3}%
-\frac{\pi }{3}\right)
\end{eqnarray*}
with 
\begin{eqnarray*}
\alpha &=&-2\Delta _{1}+\Delta _{2} \\
\beta &=&\Delta _{1}(\Delta _{1}-\Delta _{2})-\frac{1}{4}\left( \Omega
_{1}^{2}+\Omega _{2}^{2}\right) \\
\gamma &=&\frac{1}{4}\left( \Delta _{1}\Omega _{1}^{2}-\Delta _{2}\Omega
_{1}^{2}\right) \\
p &=&\sqrt{\alpha ^{2}-3\beta } \\
\cos \Theta &=&-\frac{27\gamma +2\alpha ^{3}-9\alpha \beta }{2p^{3}}.
\end{eqnarray*}
The corresponding dressed state vector for energy ${\cal E}_{\nu }$ is 
\[
\left| {\cal D}_{\nu }\right\rangle =\frac{\left( {\cal E}_{\nu }\frac{%
\Omega _{2}}{2}\left| a\right\rangle +\left( {\cal E}_{\nu }\left( {\cal E}%
_{\nu }-\Delta _{1}\right) -\frac{\Omega _{1}^{2}}{4}\right) \left|
c\right\rangle +\frac{\Omega _{1}\Omega _{2}}{4}\left| d\right\rangle
\right) }{{\cal N}_{\nu }} 
\]
with 
\[
{\cal N}_{\nu }=\sqrt{{\cal E}_{\nu }^{2}\frac{\Omega _{2}^{2}}{4}+\left( 
{\cal E}_{\nu }\left( {\cal E}_{\nu }-\Delta _{1}\right) -\frac{\Omega
_{1}^{2}}{4}\right) ^{2}+\frac{\Omega _{1}^{2}\Omega _{2}^{2}}{16}} 
\]
being the normalisation constant and $\nu =1,2,3$ indexing the dressed state.

We now consider the special case of mutual resonance of the two coupling
fields, i.e. $\Delta _{1}=\Delta _{2}=0$. In this case the dressed state
energies and vectors are 
\begin{eqnarray*}
{\cal E}_{1} &=&\frac{1}{2}\sqrt{\Omega _{1}^{2}+\Omega _{2}^{2}} \\
{\cal E}_{2} &=&0 \\
{\cal E}_{3} &=&-\frac{1}{2}\sqrt{\Omega _{1}^{2}+\Omega _{2}^{2}}
\end{eqnarray*}
and 
\begin{eqnarray*}
\left| {\cal D}_{1}\right\rangle  &=&\frac{1}{\sqrt{2}}\left( \left|
a\right\rangle +\frac{\Omega _{2}}{\sqrt{\Omega _{1}^{2}+\Omega _{2}^{2}}}%
\left| c\right\rangle +\frac{\Omega _{1}}{\sqrt{\Omega _{1}^{2}+\Omega
_{2}^{2}}}\left| d\right\rangle \right)  \\
\left| {\cal D}_{2}\right\rangle  &=&0\left| a\right\rangle -\frac{\Omega
_{1}}{\sqrt{\Omega _{1}^{2}+\Omega _{2}^{2}}}\left| c\right\rangle +\frac{%
\Omega _{2}}{\sqrt{\Omega _{1}^{2}+\Omega _{2}^{2}}}\left| d\right\rangle  \\
\left| {\cal D}_{3}\right\rangle  &=&\frac{1}{\sqrt{2}}\left( -\left|
a\right\rangle +\frac{\Omega _{2}}{\sqrt{\Omega _{1}^{2}+\Omega _{2}^{2}}}%
\left| c\right\rangle +\frac{\Omega _{1}}{\sqrt{\Omega _{1}^{2}+\Omega
_{2}^{2}}}\left| d\right\rangle \right) .
\end{eqnarray*}
These results can be used to predict the spectrum obtained when probing the
doubly driven $\left| c\right\rangle -\left| d\right\rangle $ transition via
the $\left| b\right\rangle -\left| d\right\rangle $ transition with a weak
field of frequency $\omega _{p}$ and detuning $\Delta _{p}=\omega
_{p}-\omega _{db}$. The resulting spectrum can be thought of as being made
up of a Rabi split doublet, with effective Rabi frequency $\Omega _{\text{eff%
}}=\sqrt{\Omega _{1}^{2}+\Omega _{2}^{2}},$ and a three-photon resonance
absorption peak at $\Delta _{p}=0$. The central peak is strictly only on
the three-photon resonance (i.e. satisfies $\Delta _{p}-\Delta _{1}+\Delta
_{2}=0$) when $\Delta _{1}=\Delta _{2}$, or in appropriate limits. However,
for simplicity we shall refer to the central peak as a three-photon
absorption peak when it closely approximates this resonance condition. We
note that the probe coupling will be dominated by the $\left| b\right\rangle
-\left| d\right\rangle $ transition as the other transitions are
significantly off resonance in the bare atomic basis. Also, in the strong
coupling regime ($\Omega _{1},\Omega _{2}>>\Omega _{p}$), which applies in
our experiments, almost all the population is optically pumped into the
state $\left| b\right\rangle $ with negligible population in the dressed
states. Under these conditions the probe absorption will be proportional to
the coefficient $A_{\nu }=\left| \left\langle d|{\cal D}_{\nu }\right\rangle
\right| ^{2}$. This implies that the peaks corresponding to the absorption
from state $\left| b\right\rangle $ to the outer states $\left| {\cal D}%
_{1}\right\rangle $ and $\left| {\cal D}_{3}\right\rangle $ will have equal
heights and dominate the spectrum in the limit $\Omega _{1}/\Omega _{2}>>1$,
whilst the three-photon absorption peak, corresponding to absorption to $%
\left| {\cal D}_{2}\right\rangle $, will dominate in the limit $\Omega
_{2}/\Omega _{1}>>1$. Graphs showing the energy levels ${\cal E}_{\nu }$ and
coefficients $A_{\nu }$ as a function of $\Omega _{2}/\Omega _{1}$ are
presented in Figures 2(a) and (b) respectively. These results are similar
to those presented for a ladder system by Wei {\it et al}. \cite{bib:Wei1998}

\begin{figure}[tbp]
\epsfig{figure=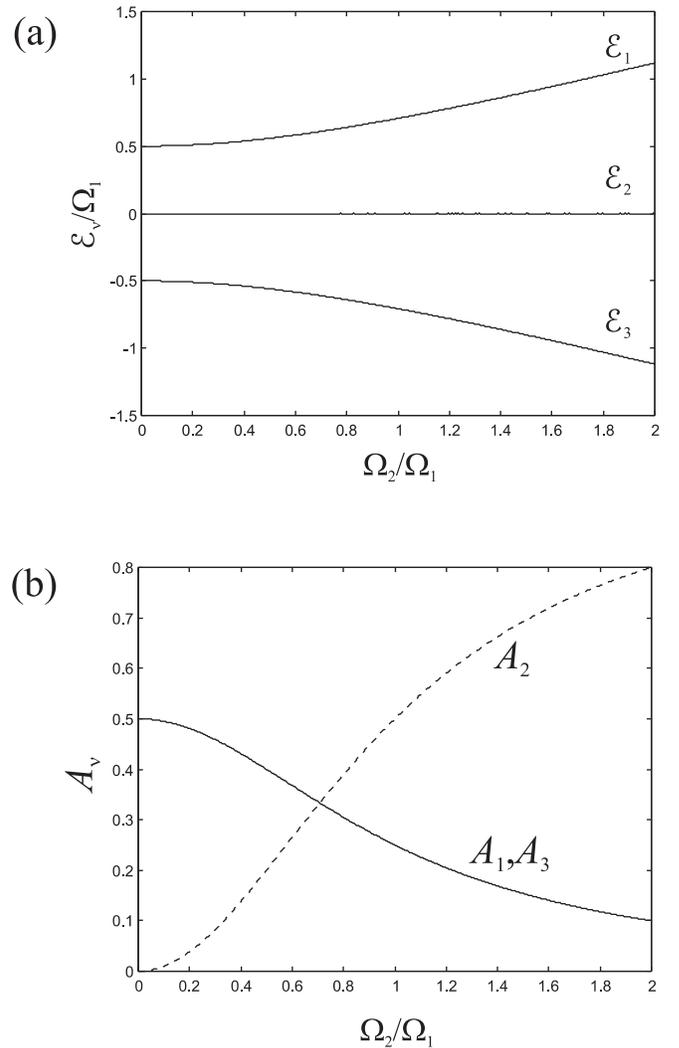,width=86mm}
\\
\caption{(a) Normalised dressed state energies, ${\cal E}_{\nu }/\Omega _{1}$ $\left(
\nu =1,2,3\right) $, as a function of the normalised Rabi frequency of the
second coupling field, $\Omega _{2}/\Omega _{1}$, with $\Delta _{1}=\Delta
_{2}=0$. (b) The coefficients $A_{\nu }=\left| \left\langle d|{\cal D}%
_{\nu }\right\rangle \right| ^{2}$ as a function of the normalised Rabi
frequency of the second coupling field, $\Omega _{2}/\Omega _{1}$.}
\end{figure}

\section{Experimental arrangement, light shifts and EIT effects}

The experimental setup is shown schematically in Figure 3. The cold $^{87}$%
Rb sample contains between $10^{7}$ and $10^{8}$ atoms in a region of
diameter approximately $3%
\mathop{\rm mm}%
$ in a standard MOT similar to the one used in our previous work on EIT\cite
{bib:EITExp}.

\begin{figure}[tbp]
\epsfig{figure=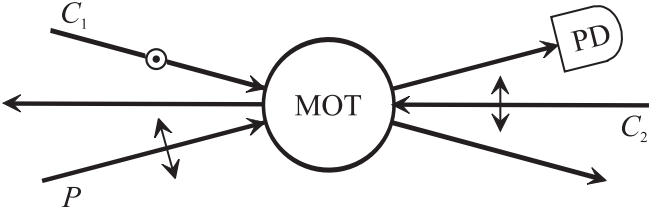,width=86mm}
\\
\caption{Schematic of the experimental arrangement showing beam polarisations.
Field $C_{1}$ is linearly polarised in the vertical plane whilst fields $%
C_{2}$ and $P$ are linearly polarised in the horizontal plane. The angle
between fields $C_{1}$ and $C_{2}$ is about $175%
{{}^\circ}%
$, whilst the angle between $C_{1}$ and $P$ is about $20%
{{}^\circ}%
$. $PD$ is an avalanche photodiode. The trapping fields are not shown
for clarity.}
\end{figure}

The laser fields are obtained from external-cavity grating-controlled diode
lasers (ECDL) in master and master-slave arrangements. The frequency of each
master laser is monitored by saturated absorption in a room temperature Rb
cell and can be locked via electronic feedback. In all master-slave
arrangements, there is an acousto-optic modulator which shifts the frequency
of the slave relative to the frequency of the master. All laser fields,
except for the second coupling field, were derived from $780%
\mathop{\rm nm}%
$ laser diodes, whilst the second coupling field was derived from a $795%
\mathop{\rm nm}%
$ laser diode.

The trapping lasers $T$ (not shown in Figure 3) are derived from a
master-slave system. They are locked and detuned by $-13%
\mathop{\rm MHz}%
$ from the $5S_{1/2}F=2$ to $5P_{3/2}F=3$ transition. The trapping beam
diameters are $\approx 1%
\mathop{\rm cm}%
$ and the total average intensity in the cold atom sample is $\approx 60%
\mathop{\rm mW}%
/%
\mathop{\rm cm}%
^{2}$.

The probe beam $P$ is derived from an ECDL and is scanned across the $%
5S_{1/2}F=2$ to $5P_{3/2}F=2$ transition by piezo-control of the external
cavity. $P$ has an average intensity $\approx 2%
\mathop{\rm mW}%
/%
\mathop{\rm cm}%
^{2}$ in a diameter $\approx 1%
\mathop{\rm mm}%
$.

The coupling beam $C_{1}$ and the trap repumping field $R$ are derived from
the same laser which is the slave of a locked ECDL. They are resonant with
the $5S_{1/2}F=1$ to $5P_{3/2}F=2$ transition. The average intensity of $%
C_{1}$ is $\approx 400%
\mathop{\rm mW}%
/%
\mathop{\rm cm}%
^{2}$ in a roughly elliptical profile $2%
\mathop{\rm mm}%
\times 4%
\mathop{\rm mm}%
$.

A second coupling beam $C_{2}$ is derived from an ECDL and is quasi-resonant
with the $5S_{1/2}F=1$ to $5P_{1/2}F=2$ transition at $795%
\mathop{\rm nm}%
$ (the D$_{1}$ line). For resonant experiments, $C_{2}$ was locked, but
for detuned experiments it was stepped using the external cavity piezo with
the frequency determined by an optical spectrum analyser. The average
intensity of $C_{2}$ can be varied up to a maximum of $\approx 300%
\mathop{\rm mW}%
/%
\mathop{\rm cm}%
^{2}$, in a beam diameter $\approx 1.3%
\mathop{\rm mm}%
$.

The probe $P$ and the second coupling field $C_{2}$ are linearly polarised
in the horizontal plane whilst the coupling field $C_{1}$ is linearly
polarised in the vertical direction. The angle between the coupling fields $%
C_{1}$and $C_{2}$ is about $175%
{{}^\circ}%
$ while the probe propagates at an angle of about $20%
{{}^\circ}%
$ with respect to $C_{1}$. These angles were found to give a good overlap of
the probe with the coupling fields in the MOT. The paths of all three
beams are coplanar.

We show in the Appendix how our arrangement of polarisations gives a good
approximation to three separate sets of N configurations of the type
depicted in Figure 1(a), together with uncoupled absorption. Although we
can qualitatively account for the uncoupled absorptions, the relative
populations of the various subsystems will be effected by the trap dynamics
and optical pumping between the schemes. Complete modelling of these effects
is beyond the scope of this work and we have therefore not presented a
quantitative theoretical estimate of the uncoupled absorption strength.

While $C_{2}$ is turned off, $C_{1}$ and $P$ form a $\Lambda $-type EIT
system. Figure 4(a) shows the probe absorption versus probe detuning with $%
C_{1}$ locked to the $5S_{1/2}F=1$ to $5P_{1/2}F=2$ transition in the
saturated absorption cell, and with $C_{2}$ turned off. The probe detuning
is taken to be zero when the probe frequency is equal to the $5S_{1/2}F=2$
to $5P_{3/2}F=2$ transition frequency in the saturated absorption cell. It
is seen that the spectrum consists of a central peak situated between the
two Autler-Townes peaks of a standard asymmetric EIT profile expected with
detuned coupling field. The detuning of $C_{1}$ is due to the fact that
the $5S_{1/2}F=1$ level of the sample is light-shifted with respect to the
same level in the saturated absorption cell because of the interaction of $%
C_{1}$ with neighbouring transitions (mainly the $5S_{1/2}F=1$ to $%
5P_{3/2}F=1$). We have estimated this light shift to be $\Delta _{1}\approx 7%
\mathop{\rm MHz}%
$. The central peak in the spectrum is caused by $P$ probing Zeeman
sublevels that are not coupled by $C_{1}$ {\bf (}as discussed in the
Appendix). We call this peak the uncoupled absorption peak; in our
previous work \cite{bib:EITExp} we were unable to resolve this peak since
the coupling field was too weak. It is to be noted that all three peaks in
Figure 4(a) are displaced by approximately $9%
\mathop{\rm MHz}%
$ with respect to the saturated absorption transition. This is because the
trapping beams $T$ act as a detuned coupling field with $P$ in a V-type EIT
configuration, as can be seen from Figure 1(b). This splits each of the
three peaks into two, one of which is very much larger than the other
because $T$ is detuned by $-13%
\mathop{\rm MHz}%
$. Of the smaller peaks, only the one corresponding to the red detuned
Autler-Townes peak is just visible in this trace. We note that the
linewidths of our spectra, which are seen to be up to three times the
natural linewidth of our Doppler-free sample, are broadened by beam profile
inhomogeneities, variations of Clebsch-Gordan coefficients between different
Zeeman transitions and the spread in the intensities of the six interfering
trapping beams in the MOT.

\begin{figure}[tb]
\epsfig{figure=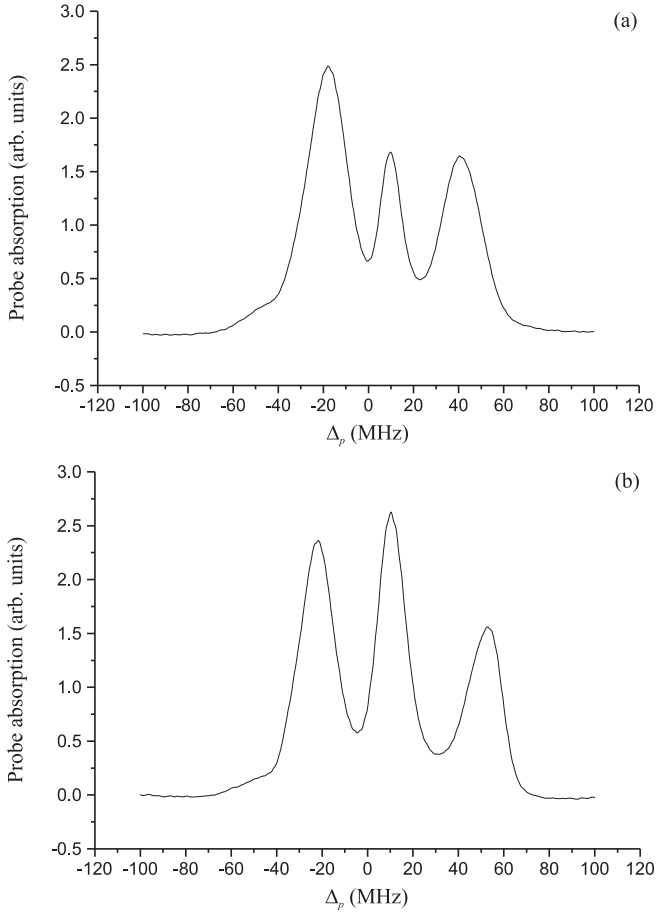,width=86mm}
\\
\caption{(a) Probe absorption spectrum with $C_{1}$ on and $C_{2}$ off. (b) Probe
absorption spectrum with both $C_{1}$ and $C_{2}$ on. Each spectrum is an
average over 200 scans.}
\end{figure}

\section{Three peak spectra of the V system}

\begin{figure}[tb]
\epsfig{figure=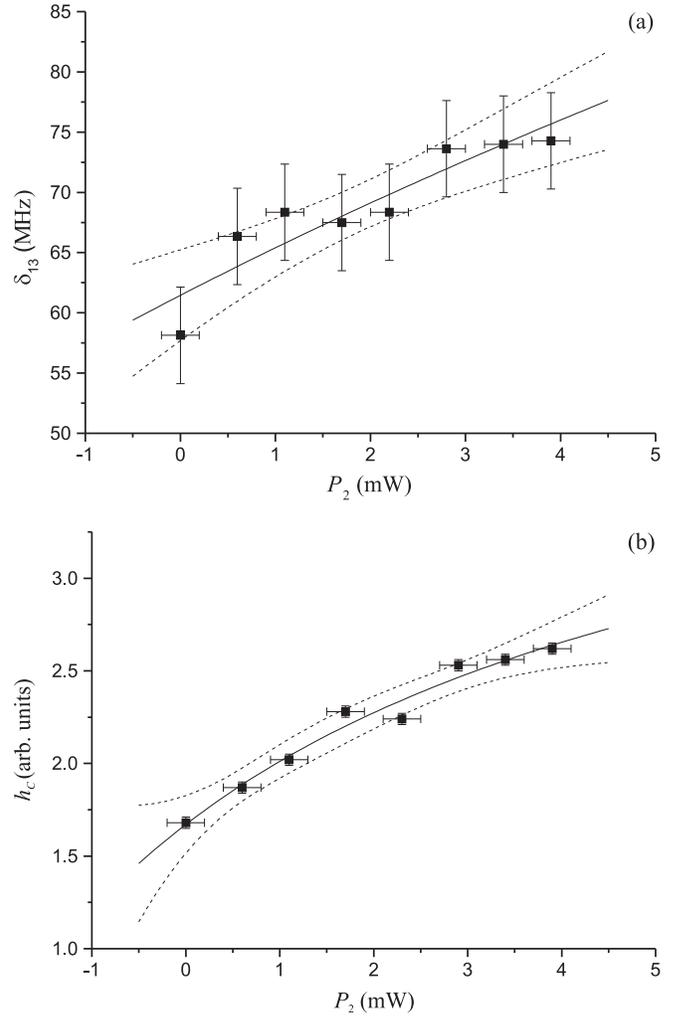,width=86mm}
\\
\caption{(a) Separation of the Autler-Townes peaks $\delta _{13}$ and (b) the height $%
h_{c}$ of the central absorption peak, plotted against the power $P_{2}$.
The solid lines are the theoretical fits with $\Omega _{1}=62\pm 2%
\mathop{\rm MHz}%
$, and $\Omega _{2}$ varying to a maximum of $\Omega _{2}=44\pm 5%
\mathop{\rm MHz}%
$ and $\Delta _{1}=\Delta _{2}\approx 7%
\mathop{\rm MHz}%
$. The dashed lines are the 95\% confidence bands.}
\end{figure}

\begin{figure}[!tb]
\epsfig{figure=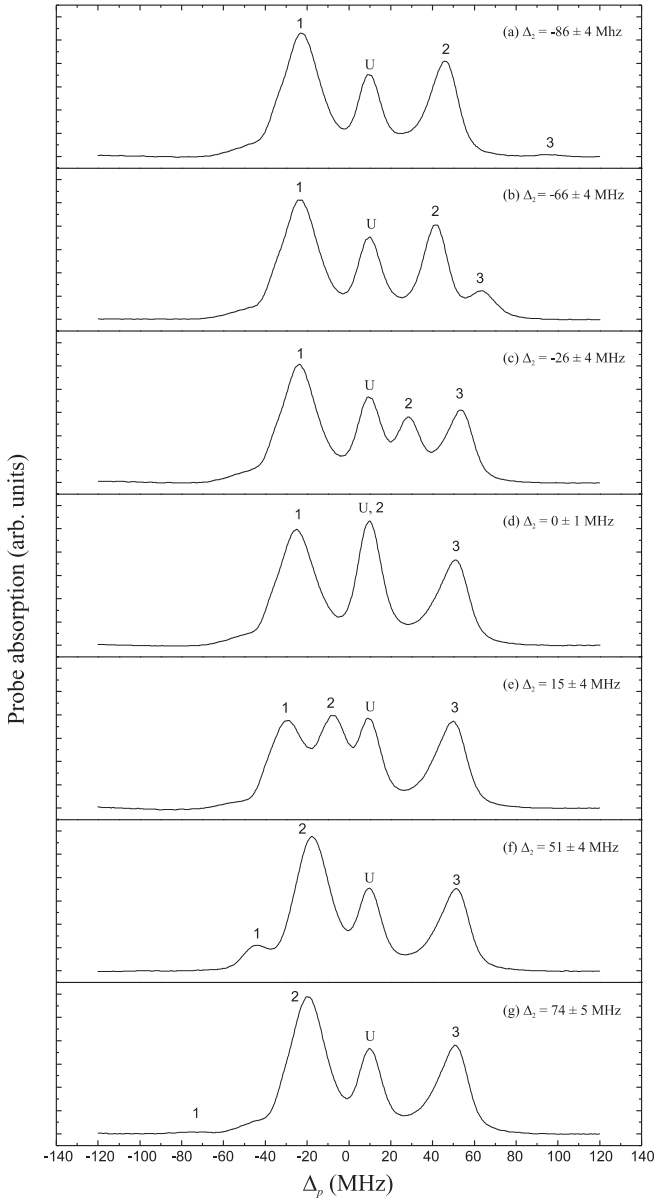,width=86mm}
\\
\caption{Probe spectra showing the migration of the three-peak structure as the
detuning $\Delta _{2}$ of the coupling field $C_{2}$ is stepped from the red
to the blue side of the resonance. The peaks are labelled 1, 2 and 3
corresponding to absorption to levels ${\cal E}_{1}$, ${\cal E}_{2}$ and $%
{\cal E}_{3}$. The uncoupled absorption peak is labelled U. Each
spectrum is an average over 200 scans.}
\end{figure}

We now describe the spectra obtained when probing the V system in the N
configuration of Figure 1. The probe absorption spectrum shown in Figure
4(b) was obtained with $C_{1}$ and $C_{2}$ tuned to their respective
transitions in saturated absorption cells; these are the same conditions
that applied in Figure 4(a) except that now $C_{2}$ is switched on and at
its maximum intensity. A comparison of the two figures shows that the
height $h_{c}$ of the central peak is larger in Figure 4(b) due to the
appearance of the three-photon absorption peak on top of the uncoupled
absorption peak, as predicted in the Section II. There is also an increase
in the splitting, $\delta _{13}={\cal E}_{1}-{\cal E}_{3}$, of the two
Autler-Townes peaks. We have taken a series of probe spectra for different
powers $P_{2}$ of $C_{2}$. The results are shown in Figures 5(a)\&(b), where
the height $h_{c}$ and the splitting $\delta _{13}$ are plotted against the
power $P_{2}$. The theoretical fits in these figures were obtained from
the theory in Section II, as follows. We assume that the contribution of
the three-photon absorption peak is proportional to $A_{2}=$ $\left|
\left\langle d|{\cal D}_{2}\right\rangle \right| ^{2},$ and that both
coupling fields have a common detuning of $\Delta =7%
\mathop{\rm MHz}%
$ from their respective transitions in the sample due to the $C_{1}$ -
induced light shift of the $5S_{1/2}F=1$ level. We find, to first order in 
$\Delta $, 
\begin{eqnarray*}
\delta _{13} &=&\sqrt{\Omega _{1}^{2}+\Omega _{2}^{2}}+{\cal O}\left( \Delta
\right) ^{2} \\
h_{c} &=&h_{uc}+B\frac{\Omega _{2}^{2}}{\Omega _{1}^{2}+\Omega _{2}^{2}}+%
{\cal O}\left( \Delta \right) ^{3}
\end{eqnarray*}
where $h_{uc}$ is the height of the uncoupled absorption peak, i.e. the
height of the central peak when $C_{2}$ is turned off, and $B$ is a
constant. These equations have been fitted to the data points and the
resulting curves shown in Figure 5(a)\&(b). The fit in Figure 5(a) gives $%
\Omega _{1}=62\pm 2%
\mathop{\rm MHz}%
$ and $\Omega _{2}=(22\pm 2%
\mathop{\rm MHz}%
/%
\mathop{\rm mW}%
^{1/2})P_{2}^{1/2}$ with a maximum of $\Omega _{2}=44\pm 5%
\mathop{\rm MHz}%
$. These Rabi frequencies are consistent with the values estimated from
the parameters of beams $C_{1}$ and $C_{2}$ and the Clebsch-Gordan
coefficients of the transitions. The fit in Figure 5(b) gives $\Omega
_{2}/\Omega _{1}=0.8\pm 0.3$ for maximum $\Omega _{2}$, which is consistent
with the previous fit; it also yields the constant $B=3\pm 1$ that
determines the relative heights of the uncoupled absorption peak and the
three-photon absorption peak. We note that for maximum power of $C_{2}$
the three-photon absorption peak accounts for approximately $0.4$ of the
total central peak height.

We now consider the case where $C_{2}$ has its maximum intensity and its
detuning is stepped across the $5S_{1/2}F=1$ to $5P_{1/2}F=2$ transition,
with $C_{1}$ tuned to the saturated absorption line. The detuning of $C_{2}$
is measured by a calibrated spectrum analyser with respect to the saturated
absorption line. The traces obtained are plotted in Figure 6. It is seen
that as $\Delta _{2}$ is stepped from the red towards the blue, the
uncoupled absorption peak, labelled $U$, remains fixed in position as
expected, but the peaks labelled 1, 2 an 3 migrate towards the red, with the
central peak 2 moving across the uncoupled absorption. The positions of
peaks 1, 2 and 3 are shown as points in Figure 7 with the curves showing the
corresponding theoretical expectations based on a detuning $\Delta _{1}=$ $7%
\mathop{\rm MHz}%
$ of $C_{1}$. We note that this behaviour is qualitatively similar to that
predicted for a ladder system in \cite{bib:Wei1998}. We note also that the
anticrossings in Figure 7 are similar to subharmonic resonances described in 
\cite{bib:PolyChro}, although because the two coupling fields are applied to
different transitions, only the first subharmonic resonances (at $\Delta
_{2}=$ $\pm \Omega _{1}/2$ in the limit of small $\Omega _{2})$ are observed
in the present study.

\begin{figure}[tbp]
\epsfig{figure=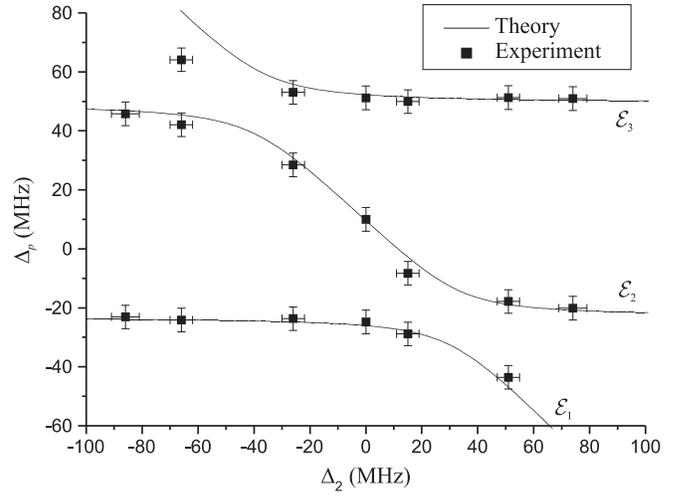,width=86mm}
\\
\caption{The positions of the absorption peaks as a function of detuning $\Delta _{2}$%
. The points are taken from the measured spectra and the curves are the
theoretical expectations based on a detuning $\Delta _{1}=$ $7%
\mathop{\rm MHz}%
$ of $C_{1}$.}
\end{figure}

\section{Conclusions}

We have presented a dressed state analysis and an experimental study of a
doubly-driven V-system probed from a fourth level in an N configuration. \
The experiments show the growth of the three-photon absorption peak and the
increasing separation of the Autler-Towns peaks as the second coupling field
intensity is increased. The migration of the three-peak probe absorption
spectrum as the detuning of one of the coupling fields is changed is also
observed. Our experiments have been carried out in a laser-cooled $^{87}$Rb
sample using the levels $5P_{3/2}F=2$, $5S_{1/2}F=1$ and $5P_{1/2}F=2$,
probed on the $5S_{1/2}F=2$ to $5P_{3/2}F=2$ transition. After taking
account of light shifts, the effects of the trapping beams and the uncoupled
absorptions in this real system, the measured spectra are in good agreement
with the analytical predictions. This investigation is important for the
understanding of a physically realisable N-system that might be used in
cross-phase modulation, photon blockade and other related studies.

\section{Acknowledgments}

We would like to thank the EPSRC for financial support on this project and
Dr T. B. Smith (Open University) for useful discussions. We would also like
to thank Roger Bence, Fraser Robertson and Robert Seaton (Open University)
for technical assistance.

\section{Appendix}

This Appendix justifies our use in Section II of a 4-state model to describe
our experiment and also shows the origin of the uncoupled absorptions. \
Figure 8(a) illustrates the individual Zeeman states of the hyperfine levels
coupled by the probe $P$ (fine lines) and two driving fields $C_{1}$ (thick
lines) and $C_{2}$ (double lines) \cite{note:SmallAngle}. All three fields
are linearly polarised with the polarisation of $C_{1}$ orthogonal to the
polarisations of $P$ and $C_{2}$ as shown in Figure 3. This particular
choice of orthogonal linear polarisations was chosen because it reduced the
uncoupled absorption of the probe field at the frequency of the $5S_{1/2}F=2$
to $5P_{3/2}F=2$ transition. Also shown are the corresponding
Clebsch-Gordan coefficients. The states $\left| m\right\rangle _{X}$ are
labeled in terms of the magnetic quantum number $m$ and $X$, where $X$ is
one of $a$, $b$, $c$ or $d$ which correspond to the $5S_{1/2}F=1$, $%
5S_{1/2}F=2$, $5P_{1/2}F=2$ and $5P_{3/2}F=2$ levels respectively in Figure
1(b).

\begin{figure}[tbp]
\epsfig{figure=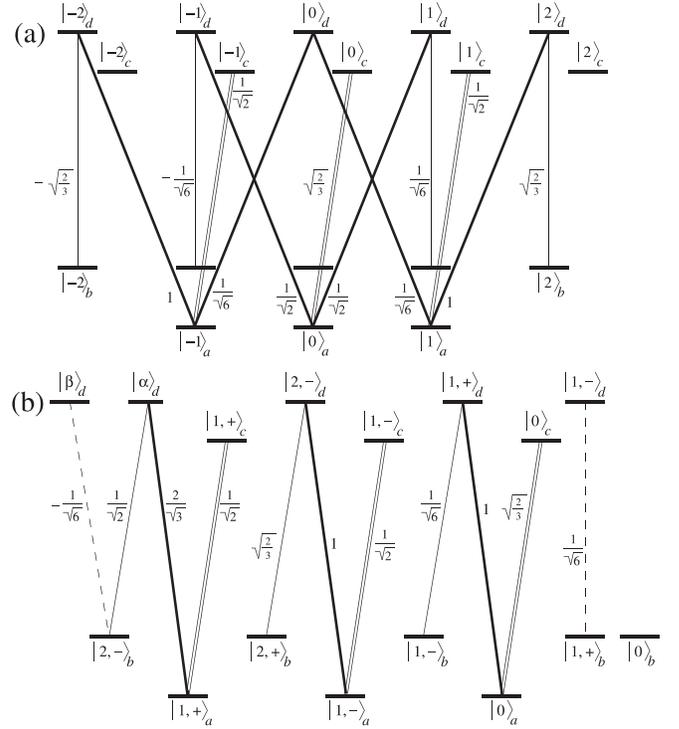,width=86mm}
\\
\caption{Zeeman states, coupled transitions and Clebsch-Gordan coefficients in the
(a) angular-momentum basis and (b) the new basis. Three separate N
configurations are clearly visible in the new basis.}
\end{figure}

The figure shows that the coupling field $C_{1}$ (thick lines) alone
provides a doubly-driven V configuration as well as a separate
quadruply-driven `W' configuration. Taking account of the additional
states coupled by the fields $C_{2}$ and $P$ it appears that a six state
and, separately, a nine state model are necessary to describe the dynamics
of our experiment. The three states, $\left| 0\right\rangle _{b}$, $\left|
-2\right\rangle _{c}$ and $\left| 2\right\rangle _{c}$ do not interact with
any fields. Surprisingly, however, this complex structure reduces to sets
of simple coupled systems with the following change of basis \cite
{note:FurtherDetails}: 
\begin{eqnarray*}
\left| m,\pm \right\rangle _{X} &\equiv &\frac{1}{\sqrt{2}}\left( \left|
m\right\rangle _{X}\pm \left| -m\right\rangle _{X}\right) \\
\left| \alpha \right\rangle _{d} &\equiv &\frac{1}{2}\left| 0\right\rangle
_{d}+\frac{\sqrt{3}}{2}\left| 2,+\right\rangle _{d} \\
\left| \beta \right\rangle _{d} &\equiv &\frac{\sqrt{3}}{2}\left|
0\right\rangle _{d}-\frac{1}{2}\left| 2,+\right\rangle _{d}.
\end{eqnarray*}
Figure 8(b) shows the transitions in this new basis. Three doubly-driven V
configurations, of the type shown in Figure 1(a), are clearly evident in the
new basis. Also shown are the effective Clebsch-Gordan coefficients for
the new transformed transitions. The dashed lines in Figure 8(b) represent
transitions probed by field $P$ but not directly coupled by either of the
coupling fields $C_{1}$ and $C_{2}$. They correspond to what we call
uncoupled absorptions of the probe.

The five coupled states on the left of Figure 8(b) shows that the
(relatively weak) field $P$ simultaneously probes a doubly-driven V
configuration as well as the $|2,-\rangle _{b}$ to $|\beta \rangle _{d}$
transition. In the steady state, the $|2,-\rangle _{b}$ to $|\beta \rangle
_{d}$ transition simply causes additional scattering of field $P$ and so
essentially this five-state system can be treated as a four-state N
configuration with an additional uncoupled absorption. The underlying probe
absorption profile is due essentially to three separate N configurations
(solid lines) with two uncoupled absorptions (dashed lines).

Each N configuration has a different set of (effective) Clebsch-Gordan
coefficients, and so the Rabi frequencies for the fields $P$, $C_{1}$ and $%
C_{2}$ differ from one N configuration to the next. Since the position of
the peaks in the probe absorption spectrum depend on the Rabi frequencies $%
C_{1}$ and $C_{2}$, the absorption peaks due to each N configuration occur
at different probe detunings. However, these differences are relatively
small, corresponding to about a 10\% shift in the separation of
Autler-Townes peaks. A {\em single} four-state model, as given in Section
II, is therefore sufficient to model our experiment, provided we allow for
broadened absorption peaks.

The square of Clebsch-Gordan coefficients of the dashed transitions is $1/6$%
. We can compare this with the different scheme in which all fields are
linear polarised along the same axis. In this case there are two uncoupled
absorption transitions from $\left| \pm 2\right\rangle _{b}$ to $\left| \pm
2\right\rangle _{d}$ for which the square of the corresponding
Clebsch-Gordan coefficients is $2/3$. That is, the absorption of the
uncoupled transitions in the parallel-linear polarisation scheme is {\em %
four times} that of the orthogonal-linear polarisation scheme for the same
degree of occupation. This clearly shows the advantage of our choice of
polarisation scheme.

In our experiment, new atoms are continuously moving into and out of the
interaction region, the trapping magnetic fields produce Zeeman mixing
amongst ground states and the trapping field weakly couples the $5S_{1/2}F=2$
level to the upper $5P_{3/2}F=3$ level. All these effects tend to
redistribute population amongst the states of the $5S_{1/2}F=2$ level. The
treatment of these effects to calculate the relative occupations of the $%
5S_{1/2}F=2$ states is, however, beyond the scope of this work. Thus, while
we can identify the transitions responsible for the uncoupled absorptions
and justify our polarisation scheme, we have no quantitative estimates of
the strength of the absorptions.

Finally we note that state $\left| 0\right\rangle _{b}$, uncoupled by any
field, is a dark state. Away from resonance conditions that produce other
dark states, all atoms would eventually decay to this state and so the
steady state would be one of complete transparency of the probe field. \
However, the above mechanisms which tend to redistribute the population of
the $5S_{1/2}F=2$ level will also tend to depopulate the state $\left|
0\right\rangle _{b}$.


\begin{references}
\bibitem{bib:EIT}  J.P. Marangos, J. Mod. Opt., {\bf 45}, 471-503 (1998); M.
Xiao, Y-Q. Li, S-Z. Jin and J. Gea-Banacloche, Phys. Rev. Lett., {\bf 74},
666-669 (1995).

\bibitem{bib:GWI}  O. Kocharovskaya, Phys. Rep. {\bf 219}, 175 (1992); M.O.
Scully, Phys. Rep. 219, 191 (1992).

\bibitem{bib:Cooling}  C. F. Roos, D. Leibfried, A. Mundt, F. Schmidt-Kaler,
J. Eschner and R. Blatt, Phys. Rev. Lett. {\bf 85}, 5547 (2000); G. Morigi,
J. Eschner, and C. Keitel, Phys. Rev. Lett. {\bf 85}, 4458 (2000).

\bibitem{bib:UltraSlow}  M.M. Kash, V.A. Sautenkov, A.S. Zibrov, L.
Hollberg, G.R. Welch, M.D. Lukin, Y. Rostovtsev, E.S. Fry and M.O. Scully,
Phys. Rev. Lett., {\bf 82}, 5229-5231 (1999); L.V. Hau, S.E. Harris, Z.
Dutton and C.H. Behroozi, Nature, {\bf 397}, 594-8 (1999) J.P. Marangos, 
{\it ibid}, {\bf 397}, 559-560 (1999).

\bibitem{bib:QND}  K.M. Gheri, P. Grangier, J-P. Poizat and D.F. Walls,
Phys. Rev. A {\bf 46}, 4276-4285 (1992).

\bibitem{bib:NLO}  S.E. Harris, J.E. Field and A. Imamoglu, Phys. Rev.
Lett., {\bf 64}, 1107-1110 (1990); G.Z. Zhang, K. Hakuta and B.P.
Stoicheff, Phys. Rev. Lett {\bf 71}, 3099 (1993); A. J. Merriam, S.J.
Sharpe, H. Xia, D. Manuszak, G.Y. Yin and S.E. Harris, Opt. Lett. {\bf 24},
625 (1999); C. Dorman, I Kucukkara and J.P. Marangos, Phys. Rev. A, {\bf 61}%
, 013802 (1999).

\bibitem{bib:PolyChro}  A.D. Greentree, C. Wei, S.A. Holmstrom, J.P.D.
Martin, N.B. Manson, K.R. Catchpole and C. Savage, J.Opt.B: Quantum
Semiclass. Opt. {\bf 1}, 240-244 (1999); A.D. Greentree, C. Wei and N.B.
Manson, Phys. Rev. A {\bf 59}, 1-4 (1999).

\bibitem{bib:Sandhya1997}  S.N. Sandhya and K.K. Sharma, Phys. Rev. A {\bf 55%
}, 2155-2158 (1997).

\bibitem{bib:Sadeghi1997}  S.M. Sadeghi, J. Meyer and H. Rastegar, Phys.
Rev. A {\bf 56}, 3097-3105 (1997).

\bibitem{bib:Wei1998}  C. Wei, D. Suter, A.S.M Windsor and N.B. Manson,
Phys. Rev. A, {\bf 58}, 2310 (1998)

\bibitem{bib:Q4WM}  M.D. Lukin, P.R. Hemmer, M. Loffler and M.O. Scully,
Phys. Rev. Lett., {\bf 81}, 2675-2678 (1998); M.D. Lukin, A.B. Matsko, M.
Fleischhauer and M.O. Scully, Phys. Rev. Lett., {\bf 82}, 1847-1850 (1999);
A.F. Huss, N. Peer, R. Lammegger, E.A. Korunsky and L. Windholz, Phys. Rev.
A {\bf 63}, 013802.

\bibitem{bib:QNLO}  A. Imamoglu, H. Schmidt, G. Woods and M. Deutch, Phys.
Rev. Lett., 79, 1467 (1997); S. Rebi\'{c}, S.M. Tan, A.S. Parkins and D.F.
Walls, J. Opt. B: Quantum Semiclass. Opt. {\bf 1}, 490 (1999); K.M. Gheri,
W. Alge, and P. Grangier, Phys. Rev. A {\bf 60}, R2673 (1999); A.D.
Greentree, J.A. Vaccaro, S.R. de Echaniz, A.V. Durrant and J.P. Marangos,
J.Opt. B: Quantum Semiclass. Opt {\bf 2} , 252-259 (2000).

\bibitem{bib:ZeemanPumping}  D.J. Fulton, S. Shepherd, R.R. Moseley, B.D.
Sinclair and M.H. Dunn, Phys.Rev.A {\bf 52}, 2302-2310 (1995).

\bibitem{bib:CPT}  H.Y. Ling, Y-Q. Li and M. Xiao, Phys. Rev. A {\bf 53},
1014-1026 (1996)

\bibitem{bib:EIA}  A.M. Akulshin, S. Barreiro and A. Lezama, Phys. Rev. A 
{\bf 57}, 2996 (1998); A. Lezama, S. Barreiro and A.M. Akulshin, Phys. Rev.
A {\bf 59}, 4732-4735 (1999); V.M. Entin, I.I. Ryabtsev, A.E. Boguslavskii
and I.M. Beterov, JETP Lett., {\bf 71}, 175-177 (2000); A.V. Taichenachev,
A.M. Tumaikin and V.I. Yudin, Phys. Rev. A {\bf 61}, 011802-1-4.

\bibitem{bib:UncoupledAbsorption}  D. McGloin, M.H. Dunn and D.J. Fulton,
Phys. Rev. A, {\bf 62}, 053802 (2000)

\bibitem{bib:Shore1990}  B. W. Shore, {\it The theory of coherent atomic
excitation, volume 2, Multilevel atoms and incoherence}, (John Wiley and
Sons, New York, 1990)

\bibitem{bib:EITExp}  S. A. Hopkings, E. Usadi, H. X. Chen and A. V.
Durrant, Opt. Comm., {\bf 138}, 185-192 (1997).

\bibitem{note:SmallAngle}  We have taken the quantization axis $z$ in Fig.
8(a) to be along the polarisation of $C_{2}$. We have also ignored a small
($\approx 7\%$ of the intensity) contribution of sigma polarisation from the
probe field due to the small ($\approx 15%
{{}^\circ}%
$) angle between the polarisation of $P$ and $C_{2}$.

\bibitem{note:FurtherDetails}  Further details of the form of the
transformation are beyond the scope of this paper and will be explored
elsewhere.
\end{references}
\end{document}